\documentclass[twocolumn,showpacs,preprintnumbers,amsmath,amssymb]{revtex4-1}

\usepackage{graphicx}
\usepackage{dcolumn}
\usepackage{bm}
\usepackage{color}
\usepackage{xcolor}

\usepackage{amsmath}
\usepackage{amssymb}
\usepackage{amsfonts}
\usepackage{bbm}
\usepackage{hyperref}

\begin{document}

\title{Realization of Qi-Wu-Zhang model in spin-orbit-coupled ultracold fermions}

\author{Ming-Cheng Liang$^{1,2}\dagger$}
\author{Yu-Dong Wei$^{1,2}\dagger$}
\author{Long Zhang$^{1,2}\dagger$}
\author{Xu-Jie Wang$^{1,2}$}
\author{Han Zhang$^{1,2}$}
\author{Wen-Wei Wang$^{1,2}$}
\author{Wei Qi$^{1,2}$}
\author{Xiong-Jun Liu$^{1,2,3*}$}
\author{Xibo Zhang$^{1,2,4}$}
\email{Corresponding authors. Emails: xiongjunliu@pku.edu.cn and xibo@pku.edu.cn.\\ $\dagger$ These authors contributed equally to this work.}
\affiliation{$^1$International Center for Quantum Materials, School of Physics, Peking University, Beijing 100871, China}
\affiliation{$^2$Collaborative Innovation Center of Quantum Matter, Beijing 100871, China}
\affiliation{$^3$Shenzhen Institute for Quantum Science and Engineering and Department of Physics, Southern University of Science and Technology, Shenzhen 518055, China}
\affiliation{$^4$Beijing Academy of Quantum Information Sciences, Beijing 100193, China}




\begin{abstract}
Based on the optical Raman lattice technique, we experimentally realize the Qi-Wu-Zhang model for quantum anomalous Hall phase in ultracold fermions with two-dimensional (2D) spin-orbit (SO) coupling. We develop a novel protocol of pump-probe quench measurement to probe, with minimal heating, the resonant spin flipping on particular quasi-momentum subspace called band-inversion surfaces. With this protocol we demonstrate the first Dirac-type 2D SO coupling in a fermionic system, and detect non-trivial band topology by observing the change of band-inversion surfaces as the two-photon detuning varies.
%
%
The non-trivial band topology is also observed by
slowly loading the atoms into optical Raman lattices and measuring the spin textures.
Our results show solid evidence for the realization of the minimal SO-coupled quantum anomalous Hall model, which can provide a feasible platform to investigate novel topological physics including the correlation effects with SO-coupled ultracold fermions. 


\end{abstract}

\pacs{}

\maketitle


Quantum anomalous Hall (QAH) effect denotes the quantum Hall effect without the Landau levels due to an external magnetic field~\cite{Xue14nsr,qi16annu}. Over three decades ago, Haldane proposed the first fundamental model for the QAH effect based on spinless fermions with staggered flux in a honeycomb lattice~\cite{Haldane88prl}. However, the QAH phase has been realized and widely studied only in the recent years~\cite{Xue13science,qah14natphys,qah14prl} in solid-state experiments based on the considerable progress of topological insulators~\cite{Kane10rmp,sczhang11rmp, Xue19review}, which has been strongly promoted by the other fundamental QAH model proposed by Qi, Wu, and Zhang based on spin-1/2 fermions in a square lattice and two-dimensional (2D) spin-orbit (SO) coupling~\cite{QiWuZhang06prb, asboth2016short}.

The Qi-Wu-Zhang model~\cite{QiWuZhang06prb} has broad impact on condensed matter research and quantum simulation. Firstly, it is the basic building block of the Bernevig-Hughes-Zhang model~\cite{bhz06science} underlying the quantum spin Hall effect~\cite{bhz06science,qsh07science,Kane10rmp,sczhang11rmp}. Secondly, it initiated a series of theoretical works~\cite{liuzhang08prl,qizhang08prb,yufang10science,nomura11prl} that inspired the successful experimental realization of the QAH effect based on magnetically doping a topological insulator~\cite{Xue13science,qah14natphys,qah14prl}. Thirdly, unlike the Haldane model where s-wave interaction cannot be directly added, the Qi-Wu-Zhang model allows for incorporating s-wave interaction and thus provides a promising route towards the realization of topological superconductor~\cite{fukane08prl,qizhang10prb,wangzhang15prb,qinglinhe17science, Kane10rmp,sczhang11rmp} and topological superfluid~\cite{liu14prl}.

Ultracold atoms provide a highly versatile platform capable of strictly implementing these 
fundamental QAH models~\cite{Esslinger15nature}. To realize the Qi-Wu-Zhang model, a novel scheme based on the optical Raman lattice technique was proposed to achieve Dirac-type 2D SO couplings in ultracold fermions~\cite{liu14prl}. Follow-up studies on 2D SO couplings have been carried out for Bose-Einstein condensates~\cite{shuai16science,liu18pra,shuai18prlrobust,shuai18prltopo}. However, the Qi-Wu-Zhang model has yet to be realized in any fermionic system. 
Such an implementation is related to the non-interacting limit of a four-Fermi-Wilson model~\cite{bermudez20arxiv,bermudez22aop} and will provide a promising platform for further studies of intriguing correlated physics in interacting regimes, including non-Abelian dynamical gauge fields~\cite{bermudez20arxiv,bermudez22aop} and topological superfluidity~\cite{liu14prl}.

Due to their insensitivity to external fields~\cite{Daley11qip}, alkaline-earth atoms (AEAs) have additional advantages for realizing highly stable SO-coupled systems and the Qi-Wu-Zhang model. AEAs enable a unique technique, the optical a.c. Stark shift  ~\cite{amohandbook96aip}, to achieve  stable and spin-dependent ground-state energy shifts~\cite{Ye08science}. Based on narrow-linewidth transitions, one-dimensional (1D) SO couplings~\cite{livi16prl, shimon17nature,Jo16pra} have been implemented, where heating due to spontaneous emission is significantly suppressed.




In this Letter, we report the first experimental realization of the Qi-Wu-Zhang model in a fermionic system.
We implement this model using strontium (${}^{87}$Sr) Fermi gases with 2D SO couplings induced by optical Raman lattices. A controlled crossover between 2D and quasi-1D SO couplings and the band topology are observed with a new protocol of pump-probe quench measurement developed here, which employs a Raman pulse to drive momentum-dependent spin-flipping. The identification of band topology is further supported by measuring the spin texture in quasi-momentum space after slowly loading the fermions into optical Raman lattices. Our work lays an experimental foundation for further studies of topological physics with ultracold fermions.

	

\begin{figure}[t]
	\includegraphics[width = 8.4cm]{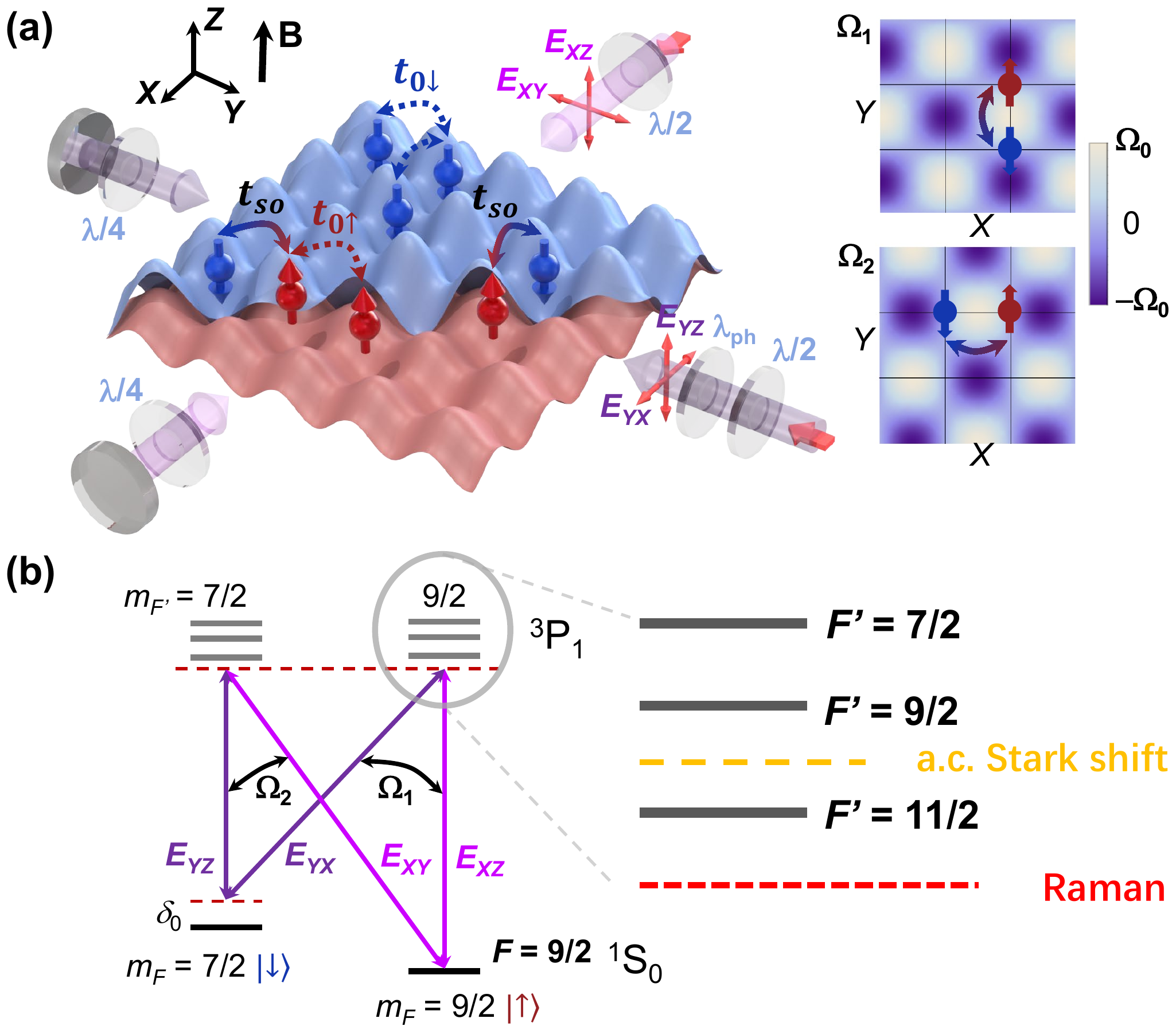}
	\caption{\label{fig1} Optical Raman lattice scheme for realizing the Qi-Wu-Zhang model and 2D SO couplings for ultracold fermions.  (a) Experimental setup. The magnetic field \textbf{B} along the $Z$ direction defines the quantization axis of atoms. Two incident lasers are reflected to construst 2D optical lattices and Raman coupling lattices. The strengths of optical lattices are different for the two spin states (red for  $|\!\!\uparrow\rangle$
		and blue for  $|\!\!\downarrow\rangle$). Two sets of Raman couplings ($\Omega_1$ and $\Omega_2$) are formed with their maximum values residing in between atomic positions (grid points). (b) Energy level diagram and the Raman coupling scheme. The relative phase $\delta \varphi$ between  $\Omega_1$ and $\Omega_2$ is controlled by a composite waveplate $\lambda_{\mathrm{ph}}$ shown in (a). }
\end{figure}

\textit{Experimental setup.---}We realize the Qi-Wu-Zhang model by implementing an optical Raman lattice scheme~\cite{liu18pra,shuai18prlrobust} in ${}^{87}$Sr Fermi gases. As shown in Fig.~\ref{fig1}(a), two  {\color{black}Raman} beams, which are both linearly polarized at a wavelength of $\lambda_0 \approx 689.4$~nm, propagate along the $\hat{X}$ and $-\hat{Y}$ horizontal directions, intersect at the atoms, and are each phase-shifted and retro-reflected to form 2D optical lattices for a pair of spin ground states: $|\!\!\uparrow\rangle \equiv$ ${}^1$S${}_0$ $|F=\frac{9}{2},m_F = \frac{9}{2}\rangle$ and $|\!\!\downarrow\rangle \equiv |\frac{9}{2}, \frac{7}{2}\rangle$. Two pairs of orthogonal polarization components, ($E_{_{XZ}}$, $E_{_{YX}}$) and ($E_{_{XY}}$, $E_{_{YZ}}$), form  two independent lattices of Raman couplings ($\Omega_1$ and $\Omega_2$) between the $|\!\!\uparrow\rangle$ and $|\!\!\downarrow\rangle$ states.  This 2D optical Raman lattice configuration  corresponds to a minimum model of QAH Hamiltonian:
\begin{eqnarray}\label{Eq:QWZHamiltonianRspace}
\hat{H} & = & \frac{\mathbf{p}^2}{2m} + V_{_\mathrm{latt}}(x,y) + \Omega_R(x,y) + \frac{\delta_0}{2}\sigma_z,
\end{eqnarray}
where $m$ is the atomic mass, $\sigma_{x,y,z}$ are Pauli matrices, and $\delta_0$ is a two-photon Raman detuning. Here, the lattice potential matrix is given by
\begin{eqnarray}
V_{_\mathrm{latt}}(x,y) & = & \left(\begin{matrix} V_{_\mathrm{latt}\uparrow}(x,y)  & 0 \\ 0 & V_{_\mathrm{latt}\downarrow}(x,y) \end{matrix}\right), \\
V_{_\mathrm{latt}\uparrow,\downarrow}(x,y) & = & V_{0X\uparrow,\downarrow}\cos^2 k_0x + V_{0Y\uparrow,\downarrow}\cos^2 k_0y, \nonumber
\end{eqnarray}
and the Raman coupling matrix
\begin{eqnarray}
\Omega_R(x,y) & = & \left(\begin{matrix} 0 & \Omega_1 + e^{i\delta\varphi}\Omega_2 \\ \Omega_1^{*} + e^{-i\delta\varphi}\Omega_2^{*}  & 0 \end{matrix} \right)
\end{eqnarray}
shall generate the SO couplings,
where $V_{0X\uparrow,\downarrow}$ ($V_{0Y\uparrow,\downarrow}$) denotes the optical lattice depth along the $X$($Y$) direction for the $|\!\!\uparrow\rangle$ or $|\!\!\downarrow\rangle$ state, $\Omega_1(x,y) = \Omega_{01}\sin k_0 x \cos k_0 y$, $\Omega_2(x,y) = \Omega_{02}\cos k_0 x \sin k_0 y$, $\delta\varphi$ is the relative phase between two sets of Raman couplings, and the lattice spacing and wavevector amplitude are given by $a = \lambda_0/2$ and $k_0 = 2\pi/\lambda_0$, respectively.
These Raman beams are detuned relative to the ${}^1$S${}_0$($F=\frac{9}{2}$)$ \rightarrow $${}^3$P${}_1$($ F'=\frac{11}{2}$) transition by -1~GHz (Fig.~\ref{fig1}(b)).
We further define an effective Zeeman splitting as $m_z \equiv \delta_0/2 + (\epsilon_{_\uparrow}-\epsilon_{_\downarrow})/2$, where $\epsilon_{_{\uparrow(\downarrow)}}$ is the onsite energy of the $|\!\!\uparrow\rangle$ ($|\!\!\downarrow\rangle$) Wannier function at $\delta_0 = 0$.
In the tight-binding regime when only the nearest-neighbor hopping is relevant, Eq.~\ref{Eq:QWZHamiltonianRspace} corresponds to the original Qi-Wu-Zhang Hamiltonian~\cite{QiWuZhang06prb,shuai16science}, $\hat{H}(\mathbf{q}) = \sum_{_{i=x,y,z}} h^i(\mathbf{q}) \sigma_i + U_{0}(\mathbf{q}) \mathbb{I}$, where $\mathbf{q}$ is the Bloch wavevector, $h^{x/y} = 2t_{_\mathrm{SO}} \sin(q_{y/x} a$), $h^z = m_z - 2\bar{t}_0(\cos(q_x a)+\cos(q_y a))$, $U_0(\mathbf{q})$ is an overall energy shift, and $\mathbb{I}$ is the identity matrix. Here, $t_{_\mathrm{SO}}$ and $\bar{t}_0=(t_{0\uparrow}+t_{0\downarrow})/2$ represent the spin-flip and mean value of spin-conserved ($t_{0\uparrow,\downarrow}$) hopping coefficients, respectively.


%
%

To isolate the $|\!\!\uparrow\rangle$ and $|\!\!\downarrow\rangle$ states from the rest of the ten nuclear spin ground states and to control their energy difference, we apply an additional ``shift beam'' to induce optical a.c. Stark shifts for  ${}^{87}$Sr atoms~\cite{SM21Liang}. As shown in Fig.~\ref{fig1}(b), the shift beam is blue-detuned by 690~MHz from the $F=\frac{9}{2} \rightarrow F'=\frac{11}{2}$ transition, which separates out an effective spin-1/2 manifold with an energy difference of about 100~kHz between  $|\!\!\uparrow\rangle$ and $|\!\!\downarrow\rangle$. Based on fractional laser intensity noise controlled to the $10^{-4}$ level, the stability of a.c. Stark shift is on the 10-Hz level, which is comparable to the ultrahigh stability in the Zeeman shift of an alkali metal atom under a 10-Gauss magnetic field with 1ppm control~\cite{shuai19rsi}.




We prepare and detect SO-coupled fermions as follows.  An almost spin-polarized ultracold ${}^{87}$Sr Fermi gas is prepared by optical pumping and subsequent evaporative cooling ~\cite{qi19cpl, SM21Liang}. About $6\times10^4$ atoms are cooled to a temperature below 200 nK; 85$\%$ of these atoms are polarized into the $|\!\!\uparrow\rangle$ state.  The shift beam intensity is ramped to its final value, and the optical Raman lattices are then turned on suddenly (for quench measurements) or slowly (for slow loading) to generate SO couplings. In the end, we shut off all lasers  within 1~$\mu$s and perform spin-resolved time-of-flight (TOF) measurements~\cite{Jo16pra,shuai18prltopo} to extract the atomic distributions $n_{\uparrow,\downarrow}$ of the $|\!\!\uparrow\rangle$ and $|\!\!\downarrow\rangle$ states~\cite{SM21Liang}. The spin texture is then given by the spin polarization $P(\mathbf{q}) = \frac{n_{\uparrow}(\mathbf{q}) - n_{\downarrow}(\mathbf{q})}{n_{\uparrow}(\mathbf{q}) + n_{\downarrow}(\mathbf{q})}$ in the first Brillouin zone.


%
%
%
%


\begin{figure}[t]
	\includegraphics[width = 8.6cm]{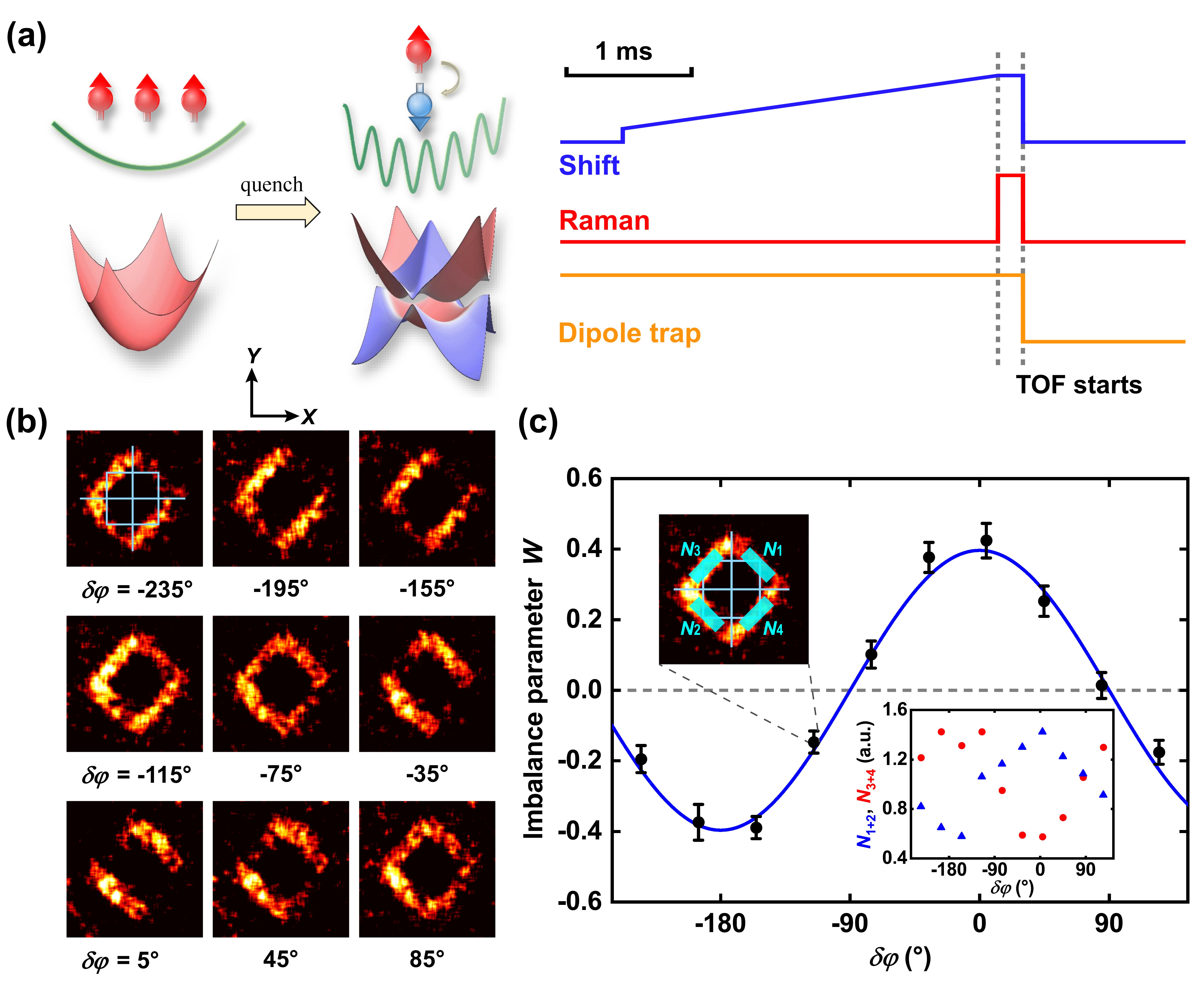}
		\caption{\label{fig2}
Pump-probe quench measurement and the demonstration of 2D SO couplings in ultracold fermions.
			%
 (a) Diagram of the pump-probe quench measurement. (b) TOF images of atoms in the $|\!\!\downarrow\rangle$ state under various relative phase $\delta\varphi$ between two Raman couplings. The cyan square marks the size of the first Brillouin zones. (c)  Crossover between 2D and quasi-1D SO couplings, presented by the imbalance parameter $W$ as a function of $\delta\varphi$. Black circles are measurements and the blue line is a fit. An optimum 2D SO coupling is achieved at $W = 0$, where the diffracted atomic population in the first and third quadrants ($N_{1+2}$) equals that in the second and fourth quadrants ($N_{3+4}$). Upper inset: illustration of $N_{1,2,3,4}$. Lower inset: original data of $N_{1+2}$ and $N_{3+4}$. Error bars represent $1\sigma$ statistical uncertainties.
		}
\end{figure}

 \textit{Pump-probe quench measurement.---}We develop a novel protocol of pump-probe quench measurement (PPQM) to probe SO couplings and band topology. As shown in Fig.~\ref{fig2}(a), we initially prepare atoms in the $|\!\!\uparrow\rangle$ state without lattice, and suddenly shine a pulse of optical Raman lattice onto these atoms for a short period of time. We then shut off all traps and perform spin-resolved TOF measurement of atomic distributions, with the results being mapped to the quasi-momentum space of the optical Raman lattice. Due to the pulse, atoms in the  $|\!\!\uparrow\rangle$ state can be selectively pumped to $|\!\!\downarrow\rangle$ at those quasi-momenta where the lowest spin-up and spin-down bands are inverted and coupled resonantly by two-photon Raman transitions, namely, at the band-inversion surface (BIS) which is a 1D ring or open line structure for the present system in 2D optical Raman lattices~\cite{zhanglin18scibull}. The BISs are an essential concept to depict  non-trivial band topology with lower-dimensional information~\cite{zhanglin18scibull,hu20prl,lee20prr,ye20pra,liu21prxqu,gong21scibull,jo19natphys,shuai19prl,du19pra}.

The PPQM technique is a new protocol that combines two important experimental methods: the pump-probe measurement, as widely applied in condensed matter experiments~\cite{jlh21jpcm,cavalieri07nature,yang21nature} and ultrafast optics studies~\cite{maiuri20jacs, zewail00acie}, and the quench measurement, as applied in previous cold-atom and solid-state experiments~\cite{greier02bnature,Mancini15science, stuhl15science,shuai18prltopo,jo19natphys,shuai19prl,sengstock16science,sengstock19natcomm,du19pra,du20prl,shen20pr,dapeng21scibull}. Compared with conventional pump-probe measurements, the PPQM protocol pumps atoms from ground state of initial Hamiltonian to the states of a completely new Hamiltonian, rather than to the excited states of the original Hamiltonian, revealing the band topology of the new Hamiltonian. Compared with conventional quench measurements, the PPQM protocol switches on the post-quench topological Hamiltonian during the application of a very short pulse ($T_{\mathrm{quench}} = 200~\mu$s~\cite{SM21Liang} in this work), which pumps the atoms to the states of the new Hamiltonian, rather than inducing oscillatory quench dynamics of a steady Hamiltonian. Thus, the PPQM technique can maximally suppress detrimental effects like heating and has the advantage in exploring intriguing topological quantum physics even with only short lifetimes. For example, the PPQM method holds the promise to promote  studies of non-Hermitian topological systems~\cite{bergholtz21rmp,zeuner15prl,zhou18science, helbig20natphys,ghatak20pnas} in the quantum regime, where the characterization of such quantum systems (e.g. those based on ultracold atoms) is often hampered by short lifetime and heating effect.

\textit{SO coupling and band topology.---}We first demonstrate a continuous crossover between 2D and quasi-1D SO couplings based on PPQM.
The relative phase $\delta\varphi$ between two Raman couplings (that are proportional to $E_{_{XZ}}E^{*}_{_{YX}}$ and $E_{_{XY}}E^{*}_{_{YZ}}$) can be tuned by a variable composite waveplate~\cite{SM21Liang} (see  $\lambda_{\mathrm{ph}}$ in Fig.~\ref{fig1}(a)) that plays the role of an electro-optic phase modulator controlling the phase shift between $E_{_{YZ}}$ and $E_{_{YX}}$~\cite{shuai18prlrobust}.
%
 %
  Fig.~\ref{fig2}(b) shows a series of momentum distribution of atoms transferred to $|\!\!\downarrow\rangle$, where the typical line segments along the $\hat{X}+\hat{Y}$ and $\hat{X}-\hat{Y}$ directions are consistent with the BIS under the corresponding experimental condition. Four groups of $|\!\!\downarrow\rangle$ atoms (marked by $N_1$ to $N_4$ in the upper inset of Fig.~\ref{fig2}(c)) appear  in accordance with the four directions, $(\pm k_0, \pm k_0)$, of SO-coupling-induced momentum transfer. As $\delta\varphi$ changes, $N_{1+2}$ shows an out-of-phase variation with respect to $N_{3+4}$ (lower inset of Fig.~\ref{fig2}(c)). We further define the population imbalance
\begin{eqnarray}
W & = & \frac{(N_1+N_2)-(N_3+N_4)}{(N_1+N_2)+(N_3+N_4)},
\end{eqnarray}
and observe that $W$ obeys a sinusoidal dependence on $\delta\varphi$
(Fig.~\ref{fig2}(c)), as shown by
%
a fit to the function $W = W_{\mathrm{max}}\cos(\delta\varphi)$~\cite{shuai18prlrobust,SM21Liang}. Here, $\delta\varphi = 0^{\circ}$ or $-180^{\circ}$ corresponds to that only $\sigma_x$ remains in the Raman coupling matrix, which is similar to 1D SO couplings for fermions in free space~\cite{jing12prl,mz12prl,Jo16pra}.
By contrast, the optimal 2D Dirac-type SO coupling~\cite{liu18pra} is achieved at $\delta\varphi = \pm 90^{\circ}$, where balanced $|\!\!\downarrow\rangle$ populations of $N_1\sim N_4$ are observed with $W=0$.
These results reveal the crossover between 2D and quasi-1D SO couplings in our fermionic system, where the
optimal 2D SO coupling is chosen as the experimental condition for subsequent measurements.

\begin{figure}[t]
	\includegraphics[width = 8.6cm]{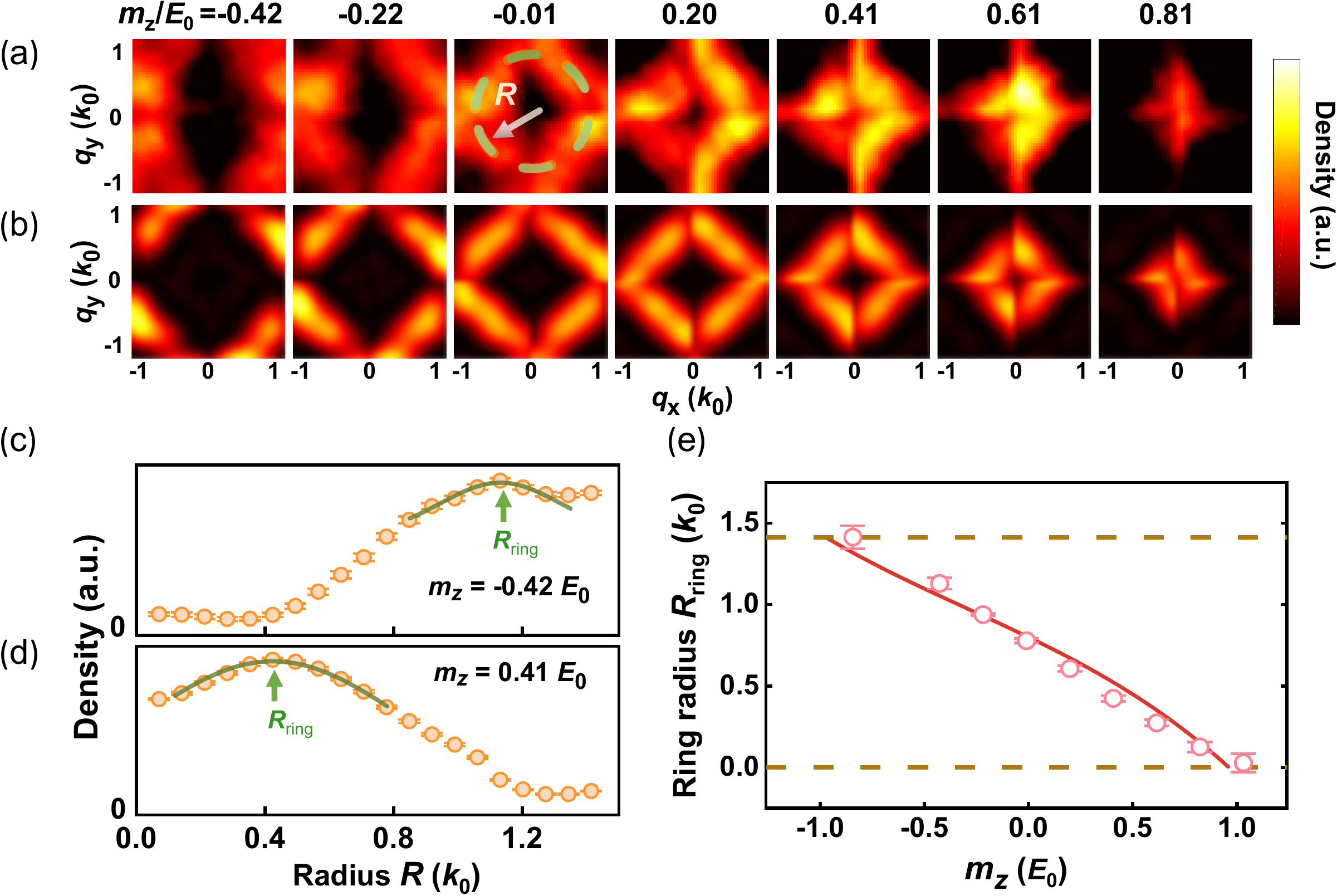}
	\caption{\label{figquench}
		Tomographic determination of post-quench band topology based on PPQM. (a)-(b) Experimental observation of band-inversion ring structures (a), represented by pumped $|\!\!\downarrow\rangle$ atoms in the first Brillouin zone after quenches, which provides a key feature of the post-quench band topology and agrees well with numerical computation (b). The distinct patterns that band-inversion rings surround the $\Gamma(0,0)$ or $M(\pm k_0, \pm k_0)$ point mark two different topological regimes. (c)-(d) Azimuthally averaged profiles of characteristic 2D distributions in (a), under $m_z/E_0 = -0.42$ (c) and $0.41$ (d), respectively, which are used to determine the ring radius $R_{\mathrm{ring}}$. (e) Measured radii $R_{\mathrm{ring}}$ as a function of $m_z$ are compared with the computed  momenta of the BIS (red solid line). Dashed lines mark $R_{\mathrm{ring}} = 0$ and $\sqrt{2}k_0$, corresponding to two boundaries of the whole topological regime and an extracted $m_z$-width of $(1.93 \pm 0.12) E_0$ based on the measurements. Error bars represent $1\sigma$ statistical uncertainties.}
\end{figure}

Next, we perform tomographic studies of the post-quench band topology based on PPQM at various two-photon detunings. We choose $V_{0X\uparrow} = V_{0Y\uparrow} = 0.6 E_0$, $V_{0X\downarrow} = V_{0Y\downarrow} = 0.3 E_0$, $\Omega_{01} = 0.53 E_0$ and $\Omega_{02} = 0.22 E_0$, where $E_0 = \frac{\hbar^2k_0^2}{2m}$ is the recoil energy. Fig.~\ref{figquench}(a) shows that under different $m_z$ values,  atoms are pumped to the $|\!\!\downarrow\rangle$ state at different quasi-momenta in the first Brillouin zone (FBZ), and the maximum density of these $|\!\!\downarrow\rangle$ atoms shows ring-like structures in the 2D distributions. As $m_z$ increases from negative to positive, the ring structure shrinks towards the center of FBZ ($\Gamma$ point), which characterizes a topological transition~\cite{zhanglin18scibull}. Fig.~\ref{figquench}(b) shows numerical simulations that are very similar to the measurements. We further perform azimuthal averaging of each 2D distribution in Fig.~\ref{figquench}(a) and then extract a ``ring radius'' $R_{\mathrm{ring}}$ that corresponds to the maximum density in the 1D profile, as showcased in Fig.~\ref{figquench}(c) and (d). In Fig.~\ref{figquench}(e), the measured ring radii $R_{\mathrm{ring}}$ are presented together with a series of computed average radii of the BISs~\cite{SM21Liang}, showing good agreement between the measured and computed values. In particular, our measurements cross with the upper and lower boundaries of $R_{\mathrm{ring}}$ at two $m_z$
values separated by $(9.3 \pm 0.6)$~kHz,
corresponding to a width of $(1.93 \pm 0.12)~E_0$.
This result agrees well with the numerically computed $m_z$-width of 1.93~$E_0$ for the topological regime based on exact diagonalization of the Hamiltonian Eq.~\ref{Eq:QWZHamiltonianRspace}~\cite{SM21Liang} and Chern number analysis~\cite{xjl13prl,shuai16science,SM21Liang}, showing the remarkable feature that the PPQM protocol reveals accurate information of the band topology 
for the Qi-Wu-Zhang model.



\begin{figure}[t]
	\includegraphics[width = 8.4cm]{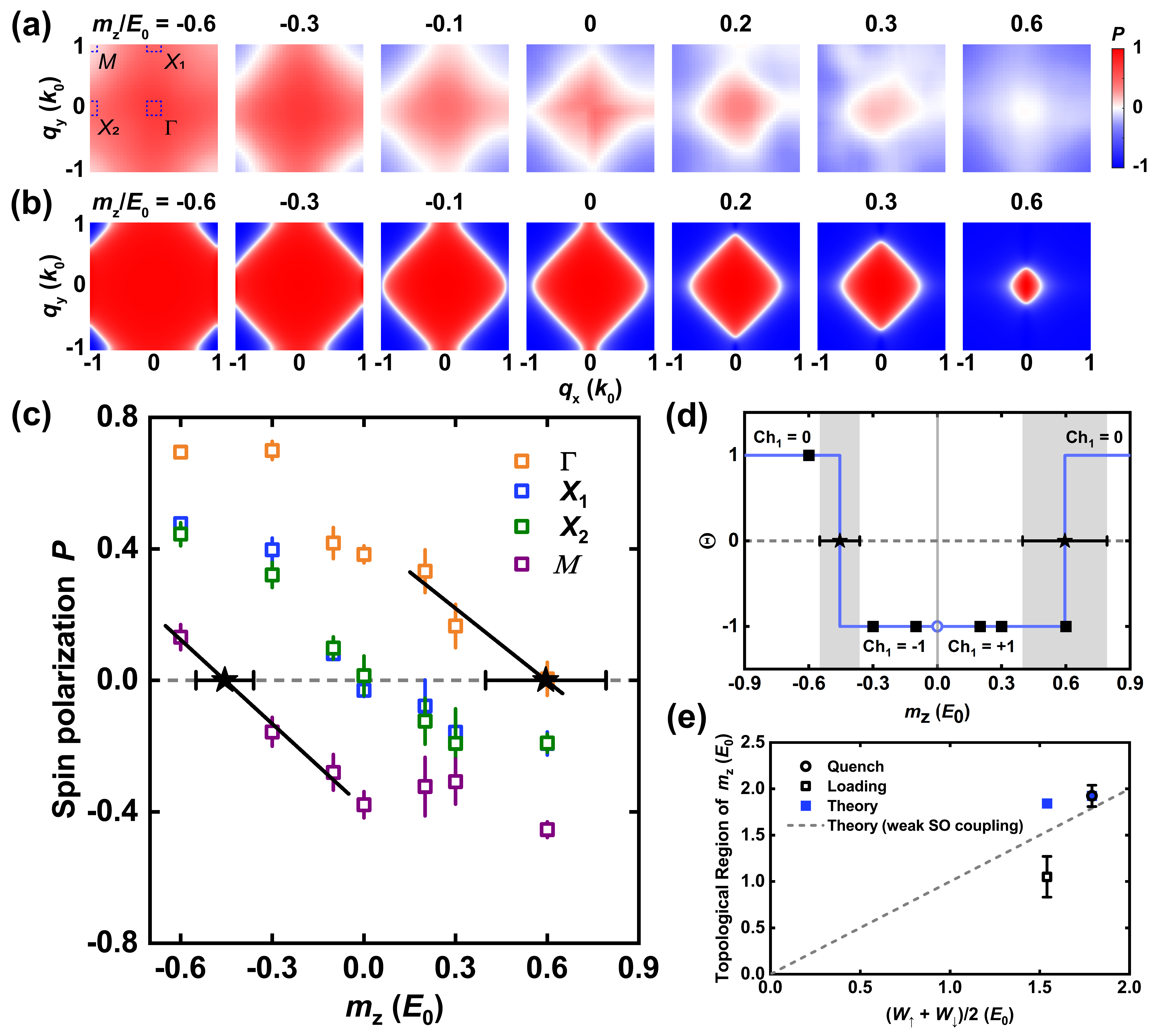}
	\caption{\label{figspintexture} Determination of band topology based on spin texture measurements.
	(a) Measured spin textures after slowly loading the atoms into optical Raman lattices. Red and blue colors denote $|\!\!\uparrow\rangle$ and $|\!\!\downarrow\rangle$, respectively. (b) Numerical simulations for zero temperature. (c) Spin polarizations at four highly symmetric momenta in the FBZ: $\Gamma(0,0)$, $X_1(0,\pm k_0)$, $X_2(\pm k_0, 0)$, and $M(\pm k_0, \pm k_0)$. (d) The sign product $\Theta = \Pi_{i=1}^4\mathrm{sgn}[P(\Lambda_i)]$ and extracted Chern number $\mathrm{Ch}_1$ as a function of $m_z$. (e) Widths of the topological regime for $m_z$: PPQM (empty circles), slow loading (empty squares), theory under experimental conditions (solid squares), and theory under vanishing SO couplings (dashed line). Here $(W_{\uparrow}+W_{\downarrow})/2$ is the average of the bare ground bandwidths for $|\!\!\uparrow\rangle$ and $|\!\!\downarrow\rangle$. Error bars represent $1\sigma$ statistical uncertainties.}
\end{figure}


\textit{Determination of band topology.---} In order to further reveal the energy band topology, we measure the spin textures after a Fermi gas is slowly loaded into the optical Raman lattices. For this purpose, the Fermi gas is initially  populated in the $|\!\!\uparrow\rangle$ state and then slowly loaded into 2D optical Raman lattices in 11 ms and further held for 1 ms. This ramp time is an order of magnitude longer than the typical inter-band relaxation time scales~\cite{SM21Liang} such that our measured spin textures reveal the property of the lowest energy band. Here, the two-photon detuning remains fixed during the loading process; we choose $V_{0X\uparrow} = V_{0Y\uparrow} = 0.7 E_0$, $V_{0X\downarrow} = V_{0Y\downarrow} = 1.2 E_0$, $\Omega_{01} = 0.80 E_0$ and $\Omega_{02} = 0.33 E_0$.
As shown in Fig.~\ref{figspintexture}(a), the majority of atoms occupy the $|\!\!\uparrow\rangle$ state at $m_z = -0.6 E_0$, while they occupy the $|\!\!\downarrow\rangle$ state at $m_z = 0.6 E_0$. The spin texture experiences a smooth change between these two cases as $m_z$ increases.
%
 Fig.~\ref{figspintexture}(b) shows the corresponding simulations for zero temperature, exhibiting similar behaviors as the measurements~\cite{SM21Liang}. Based on the spin textures, we determine the spin polarizations $P(\Lambda_i)$ at four highly symmetric points $\Lambda_{1,2,3,4} = \Gamma(0,0)$, $X_1(0,\pm k_0)$, $X_2(\pm k_0, 0)$, and $M(\pm k_0, \pm k_0)$ in the FBZ (Fig.~\ref{figspintexture}(c)), and further determine the Chern number according to the signs of $P(\Lambda_i)$~\cite{xjl13prl,shuai16science,SM21Liang}.
 As shown in Fig.~\ref{figspintexture}(d), we extract a trivial-to-topological transition at  $m_z/E_0 = (-0.46 \pm 0.09)$, and another topological-to-trivial phase boundary at  $m_z/E_0 = (0.60 \pm 0.20)$. These experimentally determined phase boundaries are to be compared with numerically computed results of $m_z/E_0 \approx \pm 0.93$; the measured topological regime has a width that is $57\%$ of the numerical result (Fig.~\ref{figspintexture}(e)). By comparison, PPQM determines a width of topological regime that is $(100\pm 6)\%$ of the numerical result.
Therefore, both the loading measurement and PPQM reveal the band topologies and are consistent with each other. Furthermore, PPQM is superior in accurately determining the phase boundaries.





\textit{Discussion and conclusion.---} We discuss the lifetime of our system.
Near $m_z = 0$, we hold the SO-coupled fermions for different periods of time, measure the decay of the total atom number, and determine a $1/e$ lifetime~\cite{SM21Liang} of $\tau_0 \gtrsim 11$~ms
in typical experimental configurations for two-spin 2D-SO-coupled Fermi gases. At present, $\tau_0$ is limited by technical impediments such as residual moving lattice potentials and has not reached the scattering-rate-limited value~\cite{SM21Liang}. In future experiments, we expect to enhance the lifetime to over $100$~ms by implementing a new optical Raman lattice scheme that eliminates moving lattice potentials. With a longer lifetime, the realization of the Qi-Wu-Zhang model in ultracold fermions shall facilitate further studies of equilibrium and non-equilibrium topological physics.

In summary, we have realized the Qi-Wu-Zhang model with 2D-SO-coupled ultracold Fermi gases. We developed a novel and robust pump-probe quench measurement protocol to probe the band topology with minimized heating effect. The band topology with 2D SO coupling is observed by measuring the BISs and spin textures. The realization of the Qi-Wu-Zhang model with spinful ultracold fermions enables the tuning of on-site interactions~\cite{Zhai20nrp} and can provide a platform for further studies of the interplay between quantum correlations and topological physics~\cite{topocorr13rev, topocorr18rev}. Future developments of our system also hold the promise to study correlated quantum dynamics~\cite{zhanglin18scibull, mcginley19prr, pastori20prr, long21prb}, simulate dynamical gauge fields~\cite{bermudez20arxiv,bermudez22aop,dalibard11rmp,goldman14rpp,wiese13ann,zohar16rpp,monika21arxiv}, and explore topological superfluids~\cite{liu14prl,qizhang10prb, cwzhang08prl,sato09prl} and topological orders in the interacting regimes~\cite{liu16njp,Galitski12prl,Cole12prl}.

\emph{Acknowledgments}.---We are grateful to Jing Zhang, Shuai Chen, Cheng Chin, Biao Wu, Jun Ye, Alejandro Bermudez, Maciej Lewenstein, Li You for insightful discussions and technical support. This work is supported by the National Key Research and Development Program of China under Grant No~2018YFA0305601, the National Natural Science Foundation of China (No.~11874073, 11825401, and 11761161003), the Open Project of Shenzhen Institute of Quantum Science and Engineering (Grant No.~SIQSE202003), the Chinese Academy of Sciences Strategic Priority Research Program under Grant No.~XDB35020100, and the Hefei National Laboratory and the Scientific and Technological Innovation
2030 under Grant No. 2021ZD0301903. X.-J.L. and X.Z. conceived the project.  M.-C.L., Y.-D.W., X.-J.W., H.Z., W.-W.W., and W.Q. performed the experiments. L.Z., M.-C.L., and Y.-D.W. performed the numerical computations. All authors contribute to the writing and revising of this manuscript.


\bibliography{FermiQWZref}


\newcommand{\pzcS}{\mathpzc{S}}
\newcommand{\pzci}{\mathpzc{i}}

\renewcommand{\theequation}{S\arabic{equation}}
\renewcommand{\thefigure}{S\arabic{figure}}
\renewcommand{\thetable}{S\arabic{table}}

\newcommand{\f}{|f\rangle}
\newcommand{\e}{|e\rangle}
\newcommand{\fth}{|f;\,+\frac{3}{2}\rangle}
\renewcommand{\eth}{|e;\,+\frac{3}{2}\rangle}
\renewcommand{\theequation}{S\arabic{equation}}
\renewcommand{\thefigure}{S\arabic{figure}}

\newcommand{\boldrho}{\mbox{\boldmath$\rho$}}
\newcommand{\bg}[1]{\mbox{\boldmath$#1$}}
\definecolor{red}{rgb}{0.7,0,0}
\definecolor{green}{rgb}{0.,0.35,0.}
\definecolor{blue}{rgb}{0.2,0.2,0.7}
\definecolor{black}{rgb}{0.15,0.15,.15}
\newcommand{\com}[1]{{\color{blue}\small\   #1 }}
\newcommand{\ad}{a^{\dagger}}
\newcommand{\al}{\alpha^{\dagger}}
\newcommand{\be}{\beta^{\dagger}}
\newcommand{\ga}{\gamma^{\dagger}}
\newcommand{\de}{\delta^{\dagger}}
\newcommand{\fd}{f^{\dagger}}
\newcommand{\an}{\mathcal N}
\newcommand{\bfn}{\mathbf n}
\newcommand{\one}{\mbox{$1 \hspace{-1.0mm}  {\bf l}$}}
\newcommand{\spec}{\mbox{spec}}
\newcommand{\bracket}{\rangle \langle}
\newcommand{\vac}{|0\rangle }
\newcommand{\ket}[1]{\left|#1\right\rangle}
\newcommand{\bra}[1]{\left\langle#1\right|}
\newcommand{\rem}[1]{\textbf{\textcolor{red}{[#1]}}}

\newcommand{\ree}{\rho_{ee}^{mm}}
\newcommand{\rgg}{\rho_{gg}^{mm}}
\newcommand{\reg}{\rho_{eg}^{mm}}
\newcommand{\rge}{\rho_{ge}^{mm}}

\setcounter{figure}{0}
\setcounter{equation}{0}
\clearpage

\section*{SUPPLEMENTAL MATERIAL}
\section{Preparation and detection of Fermi gases}
\subsection{Optical pumping and evaporative cooling}
The realization of the Qi-Wu-Zhang model for quantum anomalous Hall phase relies on two-dimensional (2D) SO couplings. In our system, this model is realized via 2D SO couplings induced by 2D optical Raman lattices based on two-photon Raman transitions between a pair of spin states of fermionic strontium ($^{87}$Sr) atoms: $\ket{\uparrow}\equiv {}^1\mathrm{S}_0 \ket{F=\frac{9}{2}, m_F=\frac{9}{2}}$ and $\ket{\downarrow}\equiv\ket{\frac{9}{2}, \frac{7}{2}}$. For $^{87}$Sr, the nuclear spin $I$ equals $\frac{9}{2}$ and the ground state $^1$S$_{0}$ has ten magnetic sublevels. Thus, it is necessary to initially polarize the atoms to the $\ket{\uparrow}$ state before loading them into optical Raman lattices.

To achieve a Fermi gas that is as spin-polarized as possible and still reach fairly low temperatures, we apply a circularly ($\sigma^{+}$-) polarized, frequency-modulated optical pumping laser
that interrogates the  ${}^1$S${}_0$$\ket{F=\frac{9}{2}} \rightarrow$ ${}^3$P${}_0$$\ket{F'=\frac{9}{2}}$ 689-nm transitions before the evaporative cooling process. 
%
%
We note that in a $100\%$-spin-polarized Fermi gas, s-wave collisions are forbidden, and p-wave (or higher-order-wave) collisions under low-temperatures are strongly suppressed by energy barriers. Therefore, 
%
%
we carefully engineer the power, frequency modulation, and pulse length of the pumping beam as well as the magnetic field such that (1) during the evaporation, atoms are partially polarized and still experience effective collisions and (2) at the end of evaporation, 
%
%
about $85\%$ of atoms occupy the state $\ket{\uparrow}$ with a temperature of less than 200~nK.

\subsection{Optical a.c. Stark shift}  
The optical a.c. Stark shift beam has a vertical polarization that is aligned with the magnetic field $\mathbf{B}$,  corresponding to a $\pi$ polarization. Due to the narrow linewidth ($\Gamma \approx 7.5$~kHz) of the ${}^1$S${}_0 \rightarrow {}^3$P${}_1$ transition and the hyperfine splitting energy structure of the ${}^3$P${}_1$ manifold, the a.c. Stark shift is non-linear with respect to the magnetic quantum number $m_F$. For example, when the differential a.c. Stark shift is 100~kHz between the $|\!\!\uparrow\rangle$  and $|\!\!\downarrow\rangle$ states, the same beam leads to a differential shift of only 75~kHz between the $|\frac{9}{2}, \frac{7}{2}\rangle$ and $|\frac{9}{2}, \frac{5}{2}\rangle$ states. Therefore, by properly choosing the frequency difference between the two Raman beams, we can realize near-resonance two-photon Raman transitions only between the $|\!\!\uparrow\rangle$ and $|\!\!\downarrow\rangle$ states; that is to say, the a.c. Stark shift beam isolates the $|\!\!\uparrow\rangle$ and $|\!\!\downarrow\rangle$ states from the rest of the spin ground states.

\subsection{Spin-resolved time-of-flight measurements}
In order to extract the momentum distributions of the $\ket{\uparrow}$ and $\ket{\downarrow}$ states, we perform a spin-resolved time-of-flight (TOF) detection that comprises of	three measurements under different conditions. 
In the first measurement,
shortly after the traps are shut off and TOF starts, we pulse on a $\sigma^{+}$-polarized, frequency-modulated `Blast' beam~\cite{Jo16pra} that removes the $\ket{\uparrow}$ state by interrogating the 
$\ket{F=\frac{9}{2},m_F=\frac{9}{2}} \rightarrow \ket{F'=\frac{11}{2},m'_F=\frac{11}{2}}$ transition, which is followed by additional TOF expansion and a final 461-nm absorption imaging, producing a momentum distribution $I_1$. 	
Similarly, in the second measurement, we change the frequency modulation range of  the Blast beam to remove both the $\ket{\uparrow}$ and $\ket{\downarrow}$ states, producing a momentum distribution $I_2$. 	In the third measurement, we apply no Blast pulse and measure the momentum distribution $I_0$ of all spin states.	Based on these three measurements, the momentum distributions of the $\ket{\uparrow}$ and $\ket{\downarrow}$ states are extracted as $I_0-I_1$ and $I_1-I_2$, respectively.

In Fig.~2 of the main text, the TOF images show the momentum distributions. In Figs.~3 and 4, we map the TOF results to the quasi-momentum space of the optical Raman lattice with a procedure similar to that employed in the works on 2D-SO-coupled ultracold bosons~\cite{shuai16science,shuai18prltopo}.

\section{Coherence between two 1D SO couplings}

The realization of a 2D SO coupling relies on coherent superposition of two lattices of Raman couplings. Here, we first use a 1D SO-coupled Fermi gas to verify the coherence between two sets of Raman couplings in the double-$\Lambda$ configuration (see Fig.~1(b) in the main text). 	In this simplified case, we can directly measure the Raman Rabi oscillation and show the variation of a total Rabi frequency as a function of a controlled relative phase between two sets of Raman couplings.




In a setup similar to that in Fig.~1(a) of the main text but without the retro-reflecting mirrors, two linearly polarized ``Raman'' beams cross at the atoms, with their polarizations each rotated by $58^{\circ}$ from the vertical direction. A variable composite waveplate $\lambda_{\mathrm{ph}}$ 
%
%
is placed in the Raman beam path along the $-\hat{Y}$ direction and plays the role of an electro-optic phase modulator that tunes the relative phase between $E_{_{YZ}}$ and $E_{_{YX}}$~\cite{shuai18prlrobust}; see section~\ref{subsec:compositewp} for details. In this way, the relative phase $\delta\varphi_{_\mathrm{1D}}$ between the two Raman couplings (that are proportional to $E_{_{XZ}}E^{*}_{_{YX}}$ and $E_{_{XY}}E^{*}_{_{YZ}}$) can be controlled. As illustrated by the inset of Fig.~S1, the total Rabi frequency $f_{_\mathrm{T}}$ is the superposition of two individual Raman couplings $f_1$ and $f_2$, which depends on the relative phase $\delta\varphi_{_\mathrm{1D}}$ between the two Raman SO couplings. Here, $f_{_\mathrm{T}}$ can be determined via Raman Rabi oscillation measurements. 

As shown in Fig.~S1, the measured $f_{_\mathrm{T}}$ varies with $\delta\varphi_{_\mathrm{1D}}$
and the measurements can be well described by a model $f_{_\mathrm{T}}=\sqrt{f_{1}^2+f_{2}^2+2f_{1}f_{2}\cos{(\delta\varphi_{_\mathrm{1D}})}}$, where
$\delta\varphi_{_\mathrm{1D}}=0$ is identified as the experimental condition where $f_{_\mathrm{T}}$ reaches its maximum.    
The fitted $f_{1}$ and $f_2$ agree with the values determined using the Raman beam powers and waists. 	
These results demonstrate the coherence between two Raman couplings, which lays the foundation of realizing 2D SO couplings.

\begin{figure}
	\centering 
	\includegraphics[scale=0.25]{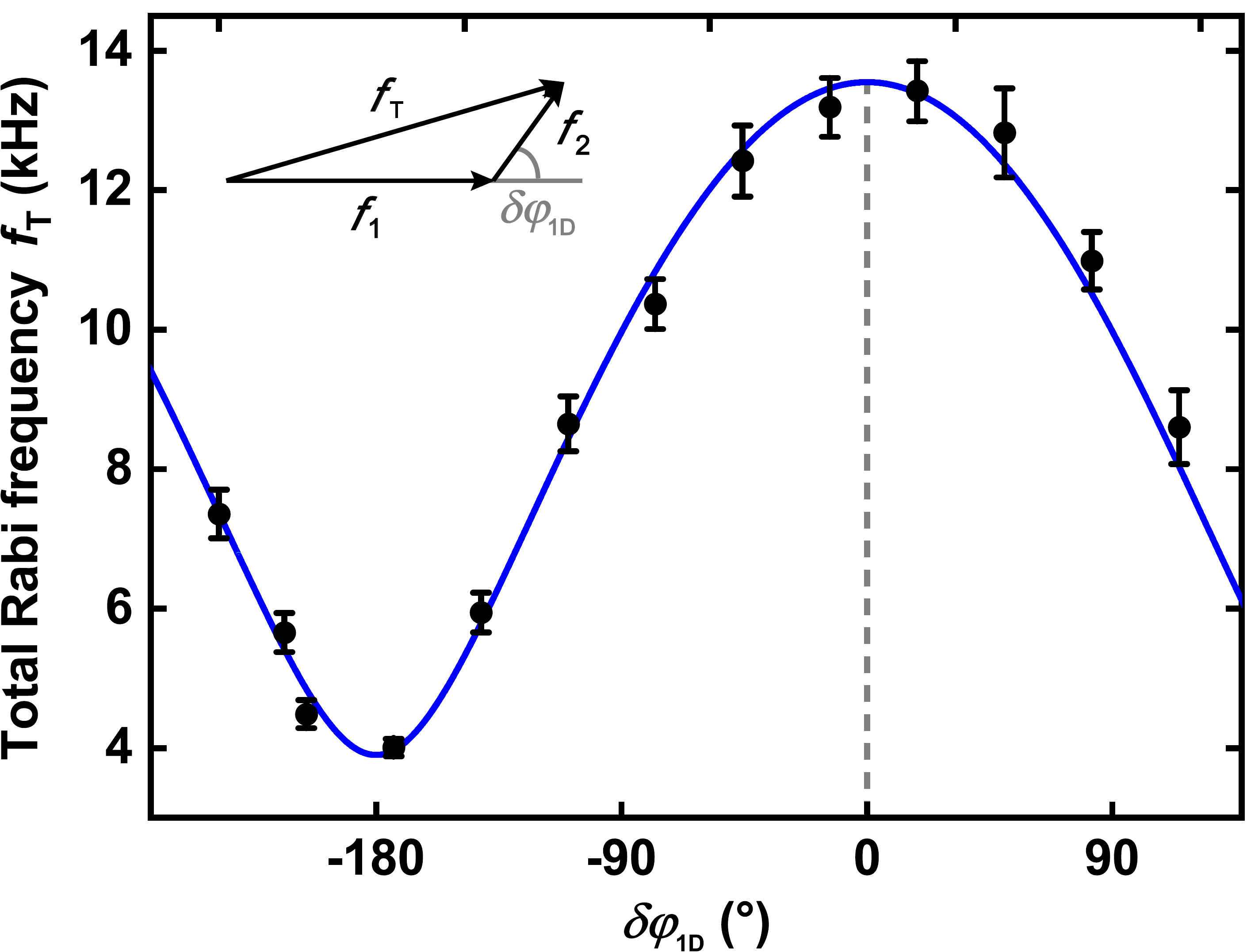}
	\caption{\label{1DTwoRamanInterference}Coherent interference between two 1D SO couplings. 
		The total Rabi frequency $f_{_\mathrm{T}}$ varies when the relative phase $\delta\varphi_{_\mathrm{1D}}$ between two Raman couplings changes. The black dots are  measurements and the blue line is a fit to a model $f_{_\mathrm{T}}=\sqrt{f_{1}^2+f_{2}^2+2f_{1}f_{2}\cos{(\delta\varphi_{_\mathrm{1D}})}}$, where the parameters are determined as $f_{1}=8.7~\mathrm{kHz}$, $f_{2}=4.8~\mathrm{kHz}$.}
\end{figure}





\section{Crossover between 2D and quasi-1D SO couplings}
Loading ultracold fermions into  2D optical Raman lattices  realizes a minimal model of quantum anomalous Hall (QAH) Hamiltonian, namely the Qi-Wu-Zhang model, as described by the Eqs.~1 to 3 in the main text.
To better understand the  crossover between 2D and quasi-1D SO coupling when $\delta\varphi$ varies,  we rewrite the Eq.~3 in the main text as follows: 
\begin{equation}\label{eq:RamanCouplingFormula}
\Omega_{R}(x,y)=\sigma_x(\Omega_{1}+\Omega_{2}\cos{\delta\varphi})-\sigma_y\Omega_{2}\sin{\delta\varphi},
\end{equation}
where $\sigma_x$ and $\sigma_y$ are Pauli matrices. 
At $\delta\varphi=\pm90^{\circ}$, $\Omega_{R}(x,y)=\sigma_x\Omega_1\mp\sigma_y\Omega_2$ represents the optimal 2D SO coupling configurations~\cite{liu18pra,shuai18prlrobust}. At  $\delta\varphi=0^{\circ} \mathrm{~or~} 180^{\circ}$, $\Omega_{R}(x,y)=\sigma_x(\Omega_1\pm\Omega_2)$ represents a configuration that closely resembles a 1D SO coupling for fermions~\cite{jing12prl,mz12prl,Jo16pra}. When $\Omega_{01} = \Omega_{02}$ and $\delta\varphi=0^{\circ} \mathrm{~or~} 180^{\circ}$, $\Omega_{R}$ will reach a purely 1D SO coupling similar to that  demonstrated in ultracold bosons~\cite{shuai18prlrobust}.

\subsection{Control of $\delta\varphi$ via a variable composite waveplate}\label{subsec:compositewp}
To tune the relative phase $\delta\varphi$ between two sets of Raman couplings, we implement a variable composite waveplate $\lambda_\mathrm{ph}$ using three waveplates. Under the basis of vertical and horizontal (V and H) polarizations, the unitary transformation matrix of a half-waveplate (HW) or a quarter-waveplate (QW) acting on a horizontally propagating  laser beam is given by
\begin{align}
U_{_\mathrm{HW}}(\theta)&=
\begin{pmatrix}
	\cos^2{\theta}-\sin^2{\theta} & 2\cos{\theta}\sin{\theta} \\ 2\cos{\theta}\sin{\theta} & \sin^2{\theta}-\cos^2{\theta}
\end{pmatrix},  \\
U_{_\mathrm{QW}}(\theta)&=
\begin{pmatrix}
	\cos^2{\theta}+i\sin^2{\theta} & (1-i)\cos{\theta}\sin{\theta} \\ (1-i)\cos{\theta}\sin{\theta} & \sin^2{\theta}+i\cos^2{\theta}
\end{pmatrix}, \notag
\end{align}
where $\theta$ represents the angle between the slow axis of a waveplate and the V direction, and the V and H polarizations corresponds to $\left(\begin{array}{c}1 \\ 0\end{array}\right)$ and $\left(\begin{array}{c}0 \\ 1\end{array}\right)$, respectively. 

In our three-waveplate setup, a half-waveplate is sandwiched between two quarter-waveplates. The slow axes of both quarter-waveplates are fixed to the same angle ($45^{\circ}$ with respect to the V direction), whereas the slow axis of the half-waveplate has a tunable angle $\varphi_{_\mathrm{HW}}$. The total transformation matrix $T_\mathrm{ph}(\varphi_{_\mathrm{HW}})$ of this variable composite waveplate $\lambda_\mathrm{ph}$ is as follows:
\begin{small}
\begin{eqnarray}
	T_\mathrm{ph}(\varphi_{_\mathrm{HW}})& = & U_{_\mathrm{QW}}(45^{\circ})U_{_\mathrm{HW}}(\varphi_{_\mathrm{HW}})U_{_\mathrm{QW}}(45^{\circ})\nonumber\\
	& = & ie^{-i\cdot2\varphi_{_\mathrm{HW}}}
	\begin{pmatrix}
		1 & 0 \\ 0 & -e^{i\cdot4\varphi_{_\mathrm{HW}}}
	\end{pmatrix}.
\end{eqnarray}
\end{small}
Therefore, by choosing $\varphi_{_\mathrm{HW}}$ for the half-waveplate, we implement an additional relative phase of $4\varphi_{_\mathrm{HW}}$ between the V and H polarizations (namely $E_{_{YZ}}$ and $E_{_{YX}}$ in the main text), which in turn tunes the relative phase $\delta\varphi$ between the two Raman couplings~\cite{shuai18prlrobust}.




\subsection{Crossover between 2D and quasi-1D SO couplings}
In our measurements for the 2D-quasi-1D crossover (Fig.~2 in the main text),  atoms are transferred from the $\ket{\uparrow}$ state to the $\ket{\downarrow}$ state by SO couplings. The Raman coupling term, Eq.~\ref{eq:RamanCouplingFormula},  can be rewritten as
\begin{small}
\begin{align}\label{fourpointsoc}
	\Omega_{R}(x,y)~\,=&~~~\,\frac{1}{4i}\left[\sigma_x(\Omega_{01}+\Omega_{02}\cos{\delta\varphi})-\sigma_y\Omega_{02}\sin{\delta\varphi}\right]\nonumber\\
	&  \times\left[e^{ik_0(x+y)}-e^{-ik_0(x+y)}\right] \notag \\
	&+\frac{1}{4i}\left[\sigma_x(\Omega_{01}-\Omega_{02}\cos{\delta\varphi})+\sigma_y\Omega_{02}\sin{\delta\varphi}\right]\nonumber\\
	&\times\left[e^{ik_0(x-y)}-e^{-ik_0(x-y)}\right].
\end{align}
\end{small}
Eq.~\ref{fourpointsoc} shows that the Raman coupling transfers momentum according to the exponential functions while it flips the spin state, which leads to the typical line segments displayed in Fig.~2(b) along the $\hat{X}+\hat{Y}$ and $\hat{X}-\hat{Y}$ directions.

In the limit of short pulses (when the pulse is short compared to the $\pi$ pulse length for a Raman Rabi frequency of $\Omega_{01}+\Omega_{02}$), the $\ket{\downarrow}$ atom numbers for momentum transfer along the $\hat{X}\pm\hat{Y}$ directions (namely $N_{1+2} = N_1 + N_2$ and $N_{3+4} = N_3+N_4$ in the main text) are proportional to  $(\Omega_{01}\pm\Omega_{02}\cos{\delta\varphi})^2+(\Omega_{02}\sin{\delta\varphi})^2$, respectively.  Thus the population imbalance is determined as follows:
\begin{align}\label{Woscillation}
W&=\frac{(N_1+N_2)-(N_3+N_4)}{(N_1+N_2)+(N_3+N_4)} \notag \\
&=\frac{2(\Omega_{01}/\Omega_{02})}{(\Omega_{01}/\Omega_{02})^2+1}\cos{\delta\varphi}.
\end{align}
Eq.~\ref{Woscillation} shows that the imbalance parameter $W$ oscillates with the relative phase $\delta\varphi$ in a form of $W=W_{\mathrm{max}}\cos{\delta\varphi}$,  where $W_{\mathrm{max}}$ is predicted to be $\frac{2(\Omega_{01}/\Omega_{02})}{(\Omega_{01}/\Omega_{02})^2+1}$ in the short-pulse limit.

In our experimental configuration, $\Omega_{01}/\Omega_{02} \approx 2.4$, which leads to a predicted $W_{\mathrm{max,theo}} \approx 0.71$ in the short-pulse limit.  We further perform numerical simulations at the actual pulse length and observe that the variation of $W$ with respect to $\delta\varphi$ is still close to a cosine function. The simulation yields an oscillation amplitude $W_{\mathrm{max, simu}} \approx 0.69$ for a 200 $\mu$s pulse length in our experiment.  From the fitted result of experimental data (Fig.~2(c) in the main text), we extract an oscillation amplitude $W_{\mathrm{max,exp}}=0.40\pm0.02$, which is smaller than the predicted value.  The difference between $W_{\mathrm{max,exp}}$ and $W_{\mathrm{max,theo}}$ or $W_{\mathrm{max,simu}}$ may be caused by the finite temperature of our Fermi gases.





\section{Pump-probe quench measurement (PPQM) and numerical simulations}
\subsection{A physical picture for PPQM}
Here we provide a simple physical picture for our  PPQM protocol. We denote the initial state as $\ket{\Psi(\mathbf{q})}=\phi(\mathbf{q})\ket{\uparrow}$, where $|\phi(\mathbf{q})|^2$ describes the atomic density at quasi-momentum $\mathbf{q}$. When the Raman lattice is pulsed on, the evolution of the quantum state is described by the evolution operator $\hat{U}(t)=\exp[-i\hat{\mathcal{H}}_{\mathrm{f}}({\bf q})t]$, where $\hat{\mathcal{H}}_{\mathrm{f}}({\bf q})$ is the post-quench Bloch Hamiltonian. Thus, the time-evolved spin polarization is determined by $\langle\sigma_z({\bf q},t)\rangle=\bra{\uparrow}\hat{U}^\dagger(t)\sigma_z\hat{U}(t)\ket{\uparrow}$. 
In other words, the spin-flip evolution is the same no matter whether atoms are prepared in an optical lattice or not. 
Furthermore, we measure the band-inversion surface~(BIS) based on the transfer of atoms from the $\ket{\,\uparrow\,}$ state to the $\ket{\,\downarrow\,}$ state, which is insensitive to the atomic distribution $|\phi({\bf q})|^2$. Therefore, our PPQM protocol not only effectively probes the BISs as has been achieved in the previous experiments~\cite{shuai18prltopo}, but also demonstrates a simplified method without the need of optical lattices before the quench, which suppresses heating and other detrimental effects.

We note that the analysis of the PPQM measurements can be further simplified under our experimental conditions.  Because we use relatively shallow lattice depths for the post-quench Hamiltonian, the ground-band atoms appear mostly in the first Brillouin zone (FBZ) after the time-of-flight. Since this work focuses primarily on the $\ket{\uparrow} \rightarrow \ket{\downarrow}$ spin flipping process between the SO-coupled ground bands, the majority of the corresponding atomic signal resides within the FBZ. We thus ignore the atomic signal outside the FBZ in the analysis of our PPQM measurements of band topology (Fig.~3 of the main text). 

\subsection{Note on the experimentally chosen pulse duration}
In the main text, we describe a short PPQM pulse 
with a duration $T_{\mathrm{quench}} = 200$~$\mu$s. This duration is chosen with the following two considerations. 

Firstly, $T_{\mathrm{quench}}$ cannot be too small. Otherwise, the signal of atoms transferred to $\ket{\downarrow}$ at the BIS will also be small and proportional to $(\Omega_{\mathrm{mod}}T_{\mathrm{quench}})^2$, where $\Omega_{\mathrm{mod}}$ is the modified Rabi frequency. This small signal from the BIS can be overwhelmed by the ``noise'' contributed by the small amount of atoms transferred to $\ket{\downarrow}$ at other quasi-momenta where $\Omega_{\mathrm{mod}}$ is higher due to the larger two-photon detuning.

Secondly, instead of studying dynamical oscillations between two spin states, we only need to pump sufficient atoms from the $\ket{\uparrow}$ state to the $\ket{\downarrow}$ state. Therefore, $T_{\mathrm{quench}}$ does not need to be too large compared with the typical period of the two-photon Raman Rabi oscillation. In fact, it only needs to reach a fraction of the duration of a $\pi$-pulse for the minimum energy gap that is reached at the BIS.

In the main text, we describe the following Raman coupling parameters used in the PPQM: $\Omega_{01} = 0.53 E_0$ and $\Omega_{02} = 0.22 E_0$, where $E_0 \approx h\times4.8$~kHz is the recoil energy at $\lambda_0 \approx 689.4$~nm, and $h$ is the Planck constant. Under an optimal 2D SO configuration, these parameters lead to a minimum energy gap (at the BIS) of $\Delta_{\mathrm{min}} \sim \sqrt{\Omega_{01}^2+\Omega_{02}^2}/2 \approx 0.29 E_0$ that also equals  the characteristic Raman Rabi frequency. The corresponding $\pi$-pulse duration can then be determined as $T_{\pi} = \frac{h}{2\Delta_{\mathrm{min}}} \approx 360$~$\mu$s. The experimentally chosen $T_{\mathrm{quench}}$ should not exceed $T_{\pi}$.

In the measurements, we have chosen $T_{\mathrm{quench}} = 200$~$\mu$s $\approx 0.56 T_{\pi}$, which satisfies the aforementioned two considerations.

\subsection{Numerical simulation of the band-inversion surfaces}
We demonstrate the effectiveness and robustness of our PPQM protocol by numerically computing the BISs based on the equilibrium spin textures. For this purpose, we employ the plane-wave basis
\begin{small}
\begin{eqnarray} \label{defwave}
	\ket{m,n}_{\uparrow}&=&C_{mn,\uparrow}\ket{q_x+2mk_0,q_y+2nk_0}, \\
	\ket{l,j}_{\downarrow}&=&C_{lj,\downarrow}\ket{q_x+2l k_0+k_0,q_y+2jk_0+k_0},  \nonumber
\end{eqnarray}
\end{small}
\noindent where $C_{mn,\uparrow}$ and $C_{lj,\downarrow}$ are coefficients for normalization.
The post-quench Bloch Hamiltonian during the Raman pulse is then expressed as
\begin{widetext}
\begin{eqnarray} \label{post-quench hamiltonian}
	\nonumber
	\hat{\cal H}_f&=&\sum_{m,n,\sigma=\uparrow,\downarrow}V_{\sigma}\left(\ket{m,n}_{\sigma}\bra{m,n+1}_{\sigma}+\ket{m,n}_{\sigma}\bra{m+1,n}_{\sigma}\right) \nonumber\\ 
	&+&\sum_{m',n'}\left[(M_x+iM_y)\ket{m',n'}_{\uparrow}\bra{m',n'}_{\downarrow}+\frac{\delta_0}{2}\ket{m',n'}_{\uparrow}\bra{m',n'}_{\uparrow}-\frac{\delta_0}{2}\ket{m',n'}_{\downarrow}\bra{m',n'}_{\downarrow}\right] \nonumber\\
	&+& \mathrm{h.c.}
\end{eqnarray}
\end{widetext}
where $M_{x/y}$ denotes the Raman couplings in the $x/y$ direction, and $\delta_0$ is the two-photon detuning.
After diagonalizing $\hat{\cal H}_f$, the BISs are identified as momenta where the $\ket{\uparrow}$ and $\ket{\downarrow}$ populations are equal, and are denoted by white lines in Fig.~S2(a) where the condition for simulations is the same as that in the post-quench Hamiltonian for Fig.~3 in the main text. 

For comparison, the simulated 2D quasi-momentum distributions of $\ket{\downarrow}$ atoms following our PPQM protocol (Fig.~3(b) in the main text) are  shown in Fig.~S2(b), which agrees well with the computed BISs. These results further support the measurement-simulation agreement shown in Fig.~3(e) of the main text and demonstrate the capability of the PPQM method to precisely measure the band topology.

\begin{figure*}[t]
\centering
\includegraphics[scale=0.6]{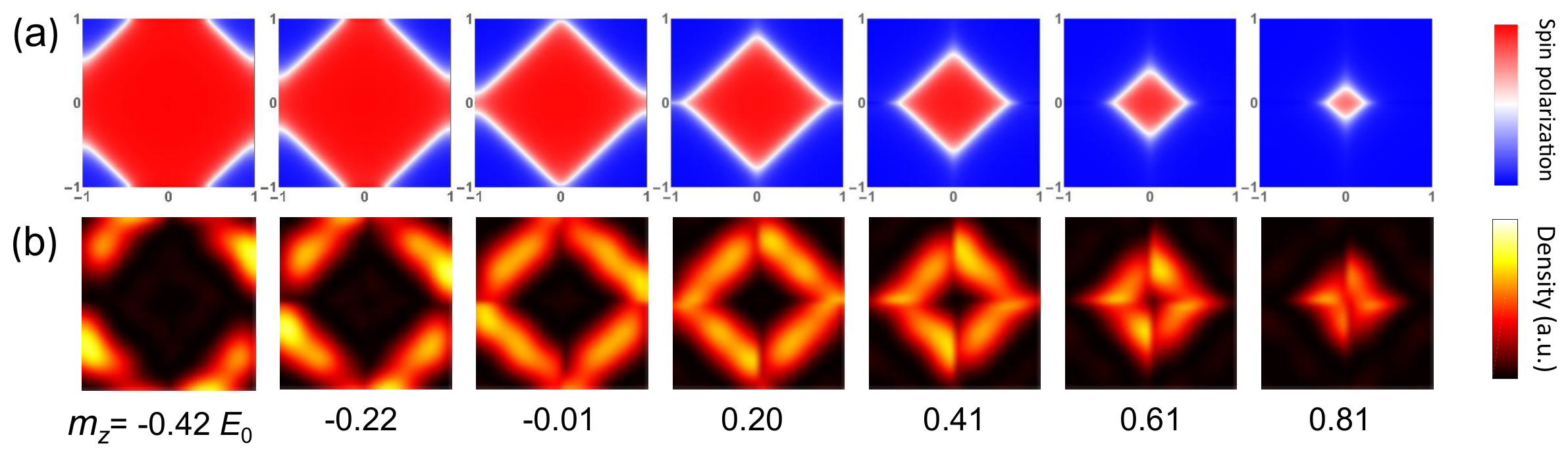}
\caption{Equilibrium spin textures and post-PPQM atomic distributions in the FBZ. (a) The computed equilibrium spin textures under the same parameters as those in the post-quench Hamiltonian for Fig.~3 in the main text. (b) Simulated distribution of $\ket{\downarrow}$ atoms in the FBZ after the PPQM. The horizontal and vertical axes correspond to the $q_x$ and $q_y$ directions, respectively.}
\label{fig:quench-diagram-2}
\end{figure*}


\section{Determine the band topology via spin-texture measurements}
The QAH Hamiltonian $\hat{H}$ (given by Eq.~1 in the main text) satisfies an inversion symmetry defined by  $\hat{P}\equiv\hat{R}_{2D}\otimes\hat{\sigma}_{z}$~\cite{xjl13prl,shuai16science}.  Here,  $\hat{R}_{2D}$ represents a 2D spatial operator that transforms the Bravais lattice vector from $\mathbf{R}$ to $-\mathbf{R}$, and $\hat{\sigma}_{z}$ is the Pauli matrix.  The Hamiltonian $\hat{H}$ can be transformed to a form expressed in the momentum space, where we obtain the Bloch Hamiltonian $\mathcal{H}(\mathbf{q})$ for a given quasi-momentum $\mathbf{q}$. It can be verified that $\hat{P}\hat{H}\hat{P}^{-1} = \hat{H}$ and that the Bloch Hamiltonian satisfies $\hat{P}\hat{\mathcal{H}}(\mathbf{q})\hat{P}^{-1}=\hat{\mathcal{H}}(-\mathbf{q})$ ~\cite{xjl13prl}. Thus at the four highly symmetric momenta $\{\Lambda_{i}\}=\{\Gamma(0,0),X_1(0,\pm k_0),X_2(\pm k_0,0),M(\pm k_0,\pm k_0)\}$ in the FBZ, we have the commutation relation $[\hat{\mathcal{H}}(\Lambda_i),\hat{P}] = 0$. Therefore, the Bloch states are eigenstates of $\hat{P}$ at the four highly symmetric momenta $\Lambda_i$ with eigenvalues $P(\Lambda_i)$, and the signs of these four eigenvalues can be used to define a topological invariant that determines the band topology~\cite{xjl13prl, shuai16science}. Here, the first Chern number is given by
\begin{eqnarray}
{\rm Ch_{1}}=-\frac{1-\Theta}{4}\sum_{i=1}^{4}{\rm sgn}[P(\Lambda_{i})],
\end{eqnarray}	
where the topological invariant $\Theta$ is defined by
\begin{eqnarray}
\Theta=\prod_{i=1}{\rm sgn}[P(\Lambda_{i})].
\end{eqnarray}	
In the experiment, measurements of spin-polarization at low but finite temperature can be applied to determine the Chern number of Bloch bands~\cite{xjl13prl,shuai16science}.

\subsection{The slow ramping-up of optical Raman lattices.} In this work, we apply a procedure similar to that used in the studies of 2D SO-coupled bosons~\cite{shuai16science,shuai18prlrobust}. Here, the Zeeman term $m_z$ is kept fixed during the ramp, while the Raman lattice beam intensities are slowly ramped up in 11~ms, such that the Fermi gas is slowly loaded into 2D optical Raman lattices. This ramp can be considered sufficiently slow based on a few considerations. Firstly, the typical energy gap at the highly symmetric points in the first Brillouin zone is about 1 kHz, corresponding to a time scale of 1 ms, which is ten times shorter than our ramp time. Secondly, even if there is some residual excitation to higher bands at the smallest energy gap, the atomic relaxation into the lowest band is also pretty fast, with a typical time scale of about 1 ms~\cite{shuai18prltopo}, which is again ten times shorter than our ramp time. Thus, the measured spin texture should be able to reflect the spin texture distributed over the lowest energy band, which provides a determination of the topology of the lowest band.

\section{Lifetime of 2D-SO-coupled Fermi gas}

\subsection{Heating rate due to optical scattering}
We estimate a heating rate by considering the single-photon scattering process. Such heating of Fermi gases in our system is mainly caused by the a.c. Stark shift beam and the optical Raman lattice beams. Here, the scattering rate is determined as
\begin{equation}
R_{\mathrm{scatt}}=\frac{\Gamma}{2}\sum_{F^{\prime}}\frac{2\Omega_{F^{\prime}}^2/\Gamma^2}{1+2\Omega_{F^{\prime}}^2/\Gamma^2+4\Delta_{F^{\prime}}^2/\Gamma^2},
\end{equation}
where $\Gamma = 7.5$~kHz is the natural linewidth of the $^1\mathrm{S}_{0}\rightarrow\,^3\mathrm{P}_{1}$ transition at 689~nm,  $\Delta_{F^{\prime}}$ and $\Omega_{F^{\prime}}$ are the frequency detuning and Rabi frequency for a single-photon transition to a manifold of excited states with $F^{\prime}=11/2$, $9/2$ and $7/2$. The corresponding heating rate is then determined by the scattering rate $R_{\mathrm{scatt}}$ and the recoil energy $E_0$:
\begin{equation}
\dot{T}=\frac{E_0 R_{\mathrm{scatt}}}{k_{\mathrm{B}}},
\end{equation}
where $k_{\mathrm{B}}$ is the Boltzmann constant.

With a relatively strong a.c. Stark shift beam (that leads to a differential shift of about 100~kHz between $\ket{\uparrow}$ and $\ket{\downarrow}$), the heating rate due to the aforementioned single-photon scattering  ($R_{\mathrm{scatt}}\sim 14$~s${}^{-1}$) by the a.c. Stark shift beam is about 3~nK/ms. We also estimate that the heating rate due to the optical Raman lattice beams is much smaller by more than a factor of ten. Thus the scattering limited lifetime of the Fermi gas is on the order of 100~ms.




\subsection{Lifetime measurement and improvement}

\begin{figure}[t]
\centering 
\includegraphics[scale=0.3]{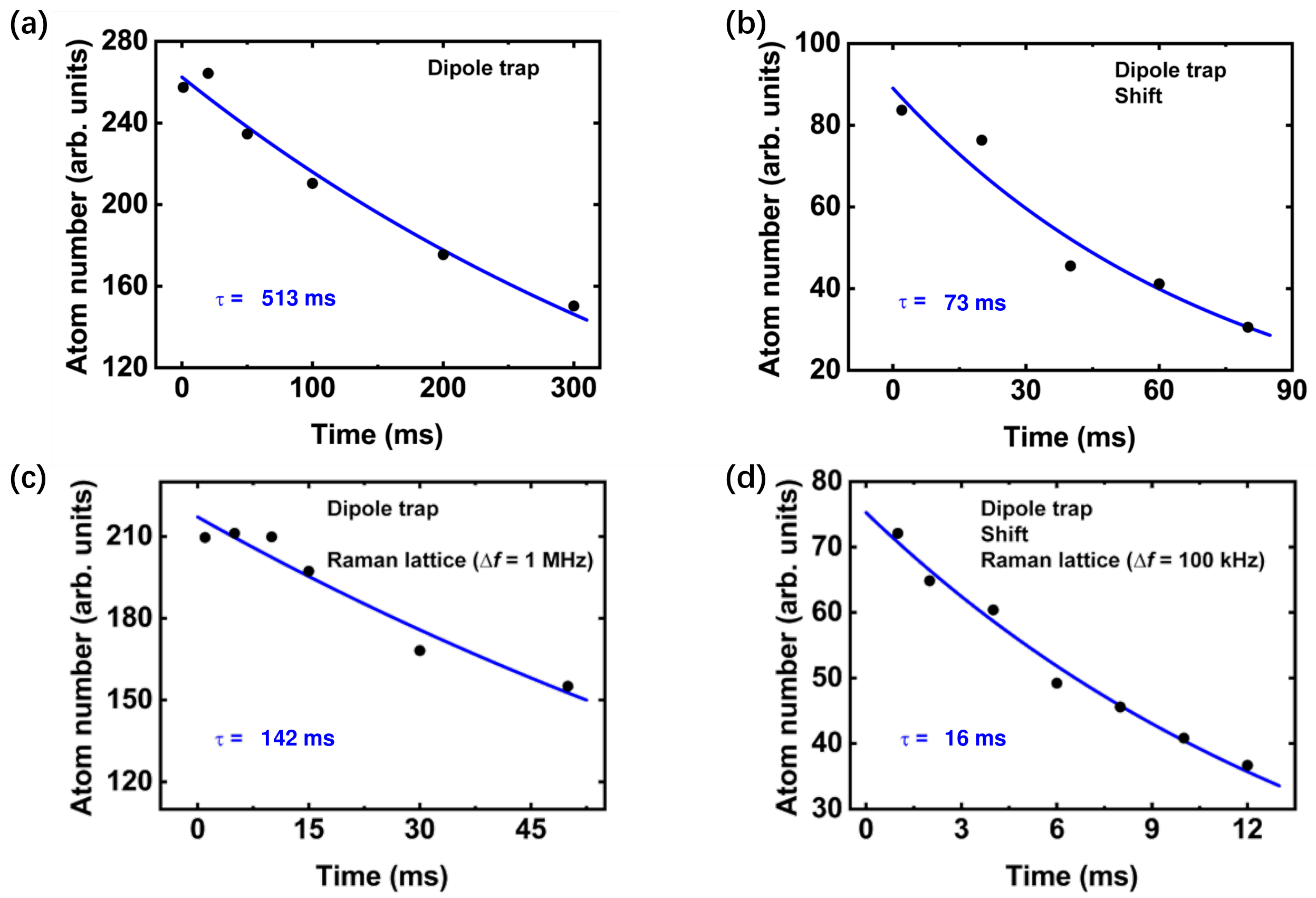}
\caption{Lifetime measurements under various experimental conditions. Black dots are experimental data, and  blue lines are exponential fits.
	%
	%
	Here, $\Delta f$ is the frequency difference between two Raman lattice beams. (a) (Measurement with) dipole trap only. (b) Dipole trap and the a.c. Stark shift beam (where the differential shift between $\ket{\uparrow}$ and $\ket{\downarrow}$ is about 100~kHz). 
	%
	(c) Dipole trap and the optical Raman lattice beams with $\Delta f = 1$~MHz (where the two-photon detuning is large). (d) Dipole trap, the a.c. Stark shift beam, and the optical Raman lattice beams with $\Delta f = 100$~kHz (where the two-photon detuning is near zero).}
\label{fig:LifetimeMeasurement}
\end{figure}

We determine the lifetime of the Fermi gas by holding the atoms under a certain experimental condition and measuring the atom number as a function of the hold time. We use
an exponential function to fit the decay of the total atom number. Under experimental conditions as used for Figs.~3 and 4 in the main text, we obtain typical $1/e$ lifetime of $\tau_{0}\approx 11 \sim 16$ ms for SO-coupled Fermi gases (see  Fig.~S3(d) for an example), which is significantly shorter than the single-photon-scattering-limited value (on the order of 100~ms). We note that $\tau_0$ is fairly insensitive to  $m_z$ when it takes values used in Figs.~3 and 4 in the main text.





As shown by Fig.~S3, the lifetime for Fermi gases with near-resonance SO couplings (Fig.~S3(d)) is significantly shorter than gases with far-off-resonance couplings (Fig.~S3(c)) or no Raman beams (Fig.~S3(b)). A likely cause of the limitation on $\tau_0$ is the heating  due to moving optical lattices~\cite{Denschlag02jpb}, which has a technical nature. Due to the limited power of the a.c. Stark shift beam (about 45~mW), The $\ket{\uparrow}$ and $\ket{\downarrow}$ states  are separated by about 100~kHz, which dictates the frequency difference between two Raman beams to be also about 100~kHz for generating near-resonance SO couplings. In our current optical design, such frequency difference between the two Raman beams (as well as their retro-reflected beams) leads to time-varying and spatially periodic  potentials, namely moving optical lattices. Such moving lattices can drag and heat the atoms, which limits the lifetime of the Fermi gas.

\noindent\textit{Discussion on improving the lifetime.---}After the completion of this work, in order to sufficiently reduce the heating effects caused by moving lattices and to enhance the atomic lifetime, we designed a new setup for the optical Raman lattices.
In this new setup, the characteristic  frequency difference underlying the moving lattices are pushed to a few MHz, which is much further away from the typical frequency scales for physical processes in our optical Raman lattice experiment. Thus the expected heating rate can be suppressed  and the lifetime can be significantly enhanced. 

Recently, we performed a preliminary test of the simplest version of the new optical setup, and observe that, encouragingly, the lifetime of 2D-SO-coupled Fermi gas has already been enhanced by more than a factor of 5. This improvement demonstrates the advantages of studying high-dimensional SO-coupled physics using AEAs, which is also one of the main motivations of the current work. A full implementation of the new optical Raman lattice design holds the promise to further increase the atomic lifetime towards the order of 100~ms. The corresponding research is well worth further investigations, which is beyond the scope of this paper and will be presented elsewhere.







\end{document}